  \documentclass[preprint]{aastex}

\received{2001 May 3}
\accepted{2001 May 18}

\slugcomment{To appear in the {\em Astronomical Journal.}}

\shorttitle{Phase Calibration Sources at $\mid b \mid \, < 2.5\arcdeg$}
\shortauthors{Wrobel, Taylor, \& Gregory}
 
\begin{document}

\title{Phase Calibration Sources in the Northern Sky\\
       at Galactic Latitudes $\mid b \mid \, < 2.5\arcdeg$}
 
\author{J.~M. Wrobel and G.~B. Taylor}
\affil{National Radio Astronomy Observatory,\\
       P.O. Box O, Socorro, New Mexico 87801, USA}
\email{jwrobel@nrao.edu,gtaylor@nrao.edu}
\and
\author{P.~C. Gregory}
\affil{Physics Department, University of British Columbia,\\
       Vancouver, British Columbia V6T 2A6, Canada}
\email{gregory@physics.ubc.ca}

\begin{abstract}
The Jodrell Bank - VLA Astrometric Survey (JVAS) of flat-spectrum
sources yielded a catalog of 2118 compact radio sources in the
northern sky.  Those sources are being used as phase calibrators for
many synthesis arrays.  JVAS suffers from a zone of avoidance because
Galactic confusion prevented selection of flat-spectrum sources at
$\mid b \mid \, < 2.5\arcdeg$.  This confusion problem was overcome by
selecting variable GB6 sources in the JVAS zone of avoidance.  A
subset of these sources were observed at 8.5~GHz with the NRAO VLA in
the JVAS style, leading to 29 compact radio sources in the zone of
avoidance whose positions have been measured to an rms accuracy of
about 10~mas or less.  This extension of the JVAS lists to lower
Galactic latitudes (1) improves the prospects for VLBA
phase-referencing of Galactic targets and (2) strengthens the lists'
usefulness for studies of the Galactic interstellar medium.
\end{abstract}


\section{MOTIVATION}

The Jodrell Bank - VLA Astrometric Survey (JVAS) of flat-spectrum
sources yielded a catalog of 2118 compact radio sources
\citep{pat92,bro98,wil98}.  Each compact source (a) has a peak flux
density at 8.4~GHz $\ge$ 50~mJy at a resolution of 200 milliarcsecond
(mas); (b) contains 80\% or more of the total source flux density; and
(c) has a position known to an rms accuracy of 12$-$55~mas.  The 2118
sources are uniformly distributed in the northern sky at Galactic
latitudes $\mid b \mid \, \ge 2.5\arcdeg$ (Figure~1).  Although these
sources are primarily intended for use as phase calibrators for the
Jodrell Bank MERLIN, they are also suitable as phase calibrators for
the NRAO Very Large Array (VLA) and can be considered as candidate
phase calibrators for the NRAO Very Long Baseline Array (VLBA) and for
millimeter arrays \citep{hol94,pec98}.

\placefigure{fig1}

JVAS suffers from a zone of avoidance because Galactic confusion
prevented selection of flat-spectrum sources at $\mid b \mid \, <
2.5\arcdeg$.  Fortunately, there is another way of selecting potential
phase calibrators that is less susceptible to Galactic confusion: seek
sources that vary with time and must, therefore, have significant
contributions from compact components.  A seven-beam receiver was used
on the former NRAO 91-m telescope, during 1986 November and 1987
October, to survey repeatedly the declination band $0\arcdeg \le
\delta_{\rm B1950} \le +75\arcdeg$ at 4.85~GHz with a resolution of
$3.5\arcmin$ \citep{con94}.  \citet{gre01} used this database
\footnote{http://pulsar.physics.ubc.ca/gregory/index.html}
to extract variability information for sources in the GB6 catalog
\citep{gre96}, the large unbiased sample of 75,162 sources derived
from the combined survey images.  GB6 sources were selected with $S
\ge 25$~mJy and with a flux density variation $\Delta S > 2.5 \sigma$
between 1986 and 1987.  The sky distribution of these long-term GB6
variables was dominantly isotropic, implying an extragalactic
population.  Imposing a further constraint of $\mid b \mid \, <
2.5\arcdeg$ on these GB6 variables leads to 97 potential phase
calibrators in the JVAS zone of avoidance.  New VLA observations in
the JVAS style of these GB6 variables would help supplement the JVAS
lists, thereby improving the prospects for VLBA phase-referencing of
Galactic targets and strengthening the lists' usefulness for studies
of the Galactic interstellar medium.

\section{OBSERVATIONS AND IMAGING}

The VLA \citep{tho80} was used 1999 June 18 UT in its A (36-km)
configuration to observe 53 of the 97 GB6 variables.  Some additional
sources initially suspected of variability were also observed
\citep{gre98}.  Data were acquired in dual circular polarizations with
bandwidths of 25~MHz and at center frequencies of 4.5351, 4.8851,
8.1149, and 8.4851~GHz.  Observations were made assuming a coordinate
equinox of 2000.  Phase calibration sources were selected from the
International Celestial Reference Frame (ICRF) as realized by VLBI
\citep{ma98}, with positions from the reference frame update of 1998-6
in the VLBA correlator database (T.M. Eubanks, private communication).
The switching time between phase calibrator observations was
6~minutes.  Additional astrometric check sources, also selected from
\citet{ma98}, were included to quantify the observed astrometric
accuracy.  Table~1 lists all phase calibrators and check sources used.

\placetable{tab1}

Each source, whether a GB6 variable, a phase calibrator, or an
astrometric check source, was scheduled in 2-minute observations.
Separate observations were needed to acquire 4.5351 and 4.8851~GHz
simultaneously, and 8.1149 and 8.4851~GHz simultaneously.  Allowing
for telescope drive and electronics settling time, this gives about 1
to 1.5~minutes of integration time per observation per source.
Additional observations of J0137$+$3309 (3C\,48), J0521$+$1638
(3C\,138), J0542$+$4951 (3C\,147), and J1331$+$3030 (3C\,286) were
used to set the flux density scale to an accuracy of 3\% and to fix
electric-vector position angles to an accuracy of $3\arcdeg$.
Instrumental polarizations were determined to an accuracy of 0.1\%
through observations of the phase calibrator sources J0359$+$5057,
J0547$+$2721, J1928$+$0848, J1935$+$2031, J2007$+$4029, and
J2322$+$5057.  Information on all Stokes parameters was obtained at
all four frequencies.  This paper tabulates astrometric and
photometric results for Stokes $I\/$ only, and at the highest possible
angular resolution at 8.4851~GHz.  Polarimetric and spectral
information at matched, and lower, angular resolutions at all four
frequencies will be presented elsewhere (T.~E. Clarke, in
preparation).  The GB6 variables were observed at {\em a priori\/}
positions taken from, in order of preference, the VLBA correlator
database, \citet{con98}, or \citet{gre96}.  All data were calibrated
using the 1999 October 15 release of the NRAO AIPS software.

The 1999 December 31 release of AIPS was used to image all phase
calibrators, astrometric check sources, and GB6 variables.  The data
were handled as described by \citet{pat92}.  Briefly, an automatic
script was employed which first located the source by making a large
image with a field of view $82\arcsec$, from which the source position
was obtained by fitting a quadratic to the brightest component using
the program MAXFIT.  If the fitted peak was less than 40~mJy per beam
area (mJy/b), then the procedure was halted and the source was
abandoned from further analysis.  This 40-mJy filter passed all phase
calibrators and check sources but only 30 of the GB6 variables.  The
automatic script next determined the source structure, by performing
two iterations of phase only self-calibration and imaging.  The final
CLEANed images of Stokes $I\/$ were convolved with a Gaussian of
200~mas FWHM and used to derive a total flux density $S_{\rm total}$,
a peak flux density $S_{\rm peak}$, and an rms noise level $\sigma$.
The error in $S_{\rm peak}$ is the quadratic sum of a 3\% amplitude
error and $\sigma$.  For one GB6 variable that passed the 40-mJy
filter, the VLA pointing position was $>40\arcsec$ from the derived
position and, because it is then difficult to distinguish source
structure from delay smearing, that source was abandoned from this
study.

Table~2 gives the astrometric and photometric results for the 29 GB6
variables that survived the automatic script and its quality control.
A code ``nvss'' in the notes for 25 GB6 variables means that the {\em
a priori\/} observed position was from \citet{con98} and the position
in Table~2 was derived from the script prior to self-calibration.  A
code ``icrf'' in the notes for 4 GB6 variables means that the {\em a
priori\/} observed position was from the VLBA correlator database and
the position in Table~2 is from that database.  For these 4 GB6
variables and the 5 astrometric check sources in Table~1, the
positions derived with the script can be compared to those from the
VLBA correlator database.  This comparison leads to a standard
deviation in right ascension of about 4~mas and in declination of
about 9~mas, which when combined in quadrature give a 2-dimensional
astrometric error of about 10~mas.  Only a few of the GB6 variables
show interesting structure at 200-mas resolution (Figure~2).

\placetable{tab2}
\placefigure{fig2}

Each compact source in the JVAS lists (a) has a peak flux density at
8.4~GHz $\ge$ 50~mJy at a resolution of 200~mas; (b) contains 80\% or
more of the total source flux density; and (c) has a position known to
an rms accuracy of 12$-$55~mas \citep{pat92,bro98,wil98}.  Each GB6
variable in Table~2 has very similar properties: it is a compact
source with (a) a peak flux density at 8.5~GHz $\ge$ 40~mJy; (b) 80\%
or more of the total source flux density; and (c) a position now known
to an rms accuracy of about 10~mas (25 sources) or less (4 sources).
These GB6 variables therefore effectively extend the JVAS lists of
compact sources into the zone of avoidance at Galactic latitudes $\mid
b \mid \, < 2.5\arcdeg$.  This has two important consequences.  First,
these GB6 sources can be considered as candidate phase calibrators for
the VLBA, thereby improving the prospects for VLBA phase-referencing
of Galactic targets.  Second, the addition of these GB6 sources to the
JVAS lists strengthens the lists' usefulness for studies of the
Galactic interstellar medium, including scintillation (eg, Cordes \&
Rickett 1998), angular broadening (eg, Desai \& Fey 2001), Faraday
rotation (eg, Wrobel 1993), and both molecular and atomic absorption
(eg, Liszt \& Lucas 1998).

\acknowledgments The authors thank Dr.\ C. Walker for comments on the
manuscript and Dr.\ J.\ Condon for help seeking NVSS positions for the
GB6 variables.  NRAO is a facility of the National Science Foundation
operated under cooperative agreement by Associated Universities, Inc.

\clearpage

\begin{deluxetable}{ll}
\tabletypesize{\scriptsize}
\tablecaption{Calibrators Used \label{tab1}} 
\tablewidth{0pc} 
\tablehead{ \colhead{ Name }& \colhead{ Name }}
\startdata 
J0102$+$5824~check& J0603$+$1742~check \\
J0228$+$6721~phase& J0725$-$0054~phase \\
J0306$+$6243~check& J1928$+$0848~phase \\
J0359$+$5057~phase& J1935$+$2031~phase \\
J0423$+$4150~phase& J1953$+$3537~phase \\
J0443$+$3441~phase& J1957$+$3338~check \\
J0547$+$2721~phase& J2007$+$4029~phase \\
J0559$+$2353~check& J2322$+$5057~phase \\
\enddata
\end{deluxetable}
\clearpage

\begin{deluxetable}{lrrrrcccl}
\tabletypesize{\scriptsize}
\tablecaption{GB6 Variable Sources at 8.5~GHz \label{tab2}} 
\tablewidth{0pc} 
\tablehead{ \colhead{                  }& 
            \colhead{ $b$              }& 
            \colhead{ $l$              }& 
            \colhead{ $\alpha_{J2000}$ }& 
            \colhead{ $\delta_{J2000}$ }& 
            \colhead{ $S_{\rm total}$  }& 
            \colhead{ $S_{\rm peak}$   }& 
            \colhead{ $\sigma$         }& 
            \colhead{ Notes            }\\
            \colhead{ Name             }& 
            \colhead{ ($\arcdeg$)      }& 
            \colhead{ ($\arcdeg$)      }& 
            \colhead{ (h,m,s)          }& 
            \colhead{ ($\arcdeg,\arcmin,\arcsec$)}& 
            \colhead{ (mJy)            }& 
            \colhead{ (mJy/b)          }& 
            \colhead{ (mJy/b)          }& 
            \colhead{                  }}
\startdata 
J0008$+$6024& $-$2.0& 117.6& 00 08 19.0524& $+$60 25 01.254&  52&  44& 0.18& nvss \\
J0223$+$6307& $+$2.1& 133.1& 02 23 29.6078& $+$63 07 17.298& 106&  98& 0.17& nvss \\
J0231$+$6250& $+$2.2& 134.1& 02 31 59.1568& $+$62 50 34.202& 125& 122& 0.16& nvss \\
J0235$+$6216& $+$1.8& 134.7& 02 35 20.6398& $+$62 16 02.334&  76&  74& 0.17& nvss \\
J0244$+$6228& $+$2.4& 135.6& 02 44 57.6969& $+$62 28 06.515& 765& 751& 0.22& icrf \\
J0317$+$5644& $-$0.6& 142.0& 03 17 54.9423& $+$56 43 57.795& 147& 143& 0.18& nvss \\
J0340$+$5405& $-$1.0& 146.1& 03 40 06.4897& $+$54 05 38.771&  50&  53& 0.16& nvss \\
J0346$+$5400& $-$0.5& 146.9& 03 46 34.5031& $+$54 00 59.100& 449& 443& 0.21& nvss \\
J0350$+$5138& $-$2.0& 148.8& 03 50 25.0508& $+$51 38 38.721& 118& 117& 0.15& nvss \\
J0438$+$4848& $+$1.3& 156.4& 04 38 59.1064& $+$48 48 46.628& 335& 308& 0.19& nvss \\
J0450$+$4056& $-$2.2& 163.7& 04 50 43.6790& $+$40 56 13.986& 367& 376& 0.23& nvss \\
J0502$+$3849& $-$1.8& 166.8& 05 02 32.4916& $+$38 49 54.937& 212& 199& 0.17& nvss \\
J0502$+$4139& $+$0.0& 164.6& 05 02 37.9876& $+$41 39 19.343& 795& 792& 0.24& nvss \\
J0541$+$3301& $+$1.4& 176.1& 05 41 49.4341& $+$33 01 31.909&  89&  93& 0.16& nvss \\
J1844$+$0137& $+$2.2&  33.6& 18 44 50.9635& $+$01 37 17.548&  52&  62& 0.19& nvss \\
J1847$+$0154& $+$1.7&  34.2& 18 47 43.9620& $+$01 54 35.500& 154& 127& 0.19& nvss \\
J1851$+$0035& $+$0.2&  33.5& 18 51 46.7231& $+$00 35 32.348& 972& 872& 0.39& nvss \\
J1855$+$0250\tablenotemark{a}&
              $+$0.4&  35.9& 18 55 35.4366& $+$02 51 19.548& 202& 189& 0.20& nvss \\
J1856$+$0610& $+$1.7&  39.0& 18 56 31.8384& $+$06 10 16.757& 311& 303& 0.20& nvss \\
J1931$+$2243& $+$1.9&  57.6& 19 31 24.9168& $+$22 43 31.259& 406& 398& 0.16& icrf \\
J1934$+$1732& $-$1.3&  53.4& 19 34 50.2056& $+$17 32 14.154& 121& 103& 0.20& nvss \\
J1936$+$2246& $+$0.9&  58.2& 19 36 29.3042& $+$22 46 25.859& 119& 111& 0.18& nvss \\
J1946$+$2300& $-$0.9&  59.5& 19 46 06.2515& $+$23 00 04.415& 201& 191& 0.18& icrf \\
J1949$+$2421& $-$0.9&  61.1& 19 49 33.1432& $+$24 21 18.252& 116& 124& 0.18& nvss \\
J2003$+$3034& $-$0.3&  68.0& 20 03 30.2448& $+$30 34 30.779& 387& 380& 0.18& nvss \\
J2028$+$3833\tablenotemark{a}&
              $-$0.2&  77.5& 20 28 54.0288& $+$38 32 48.144&  42&  49& 0.16& nvss \\
J2102$+$4702& $+$0.3&  87.9& 21 02 17.0563& $+$47 02 16.254& 170& 159& 0.15& icrf \\
J2114$+$4634& $-$1.5&  89.0& 21 14 32.8774& $+$46 34 39.296&  78&  74& 0.16& nvss \\
J2254$+$6209& $+$2.3& 109.7& 22 54 25.2918& $+$62 09 38.727&  86&  76& 0.16& nvss \\
\enddata
\tablenotetext{a}{Initial suspicion of long-term variability 
                  was not confirmed \citep{gre01}.}
\end{deluxetable}

\clearpage

\figcaption{Sky plot for 2118 compact radio sources from JVAS (small
dots) plus 29 GB6 variable sources from this work (big dots) in the
JVAS zone of avoidance.  Grey dotted line shows $b=0\arcdeg$.
\label{fig1}}

\figcaption{VLA images of Stokes $I\/$ emission at 8.3~GHz from 4
resolved GB6 variables.  Boxed circle shows the Gaussian restoring
beam at 200-mas FWHM.  Negative contours are dashed and positive ones
are solid.  These images were made from data at 8.1 and 8.5~GHz,
improving upon the script images described in Table~2.
{\em Upper left:}  J1847$+$0154.
{\em Upper right:} J1851$+$0035.
{\em Lower left:}  J1934$+$1732.
{\em Lower right:} J1946$+$2300.
\label{fig2}}

\newpage
  \epsscale{1.0} \plotone{Wrobel.fig1.eps}

\newpage 
  \epsscale{1.0} \plotone{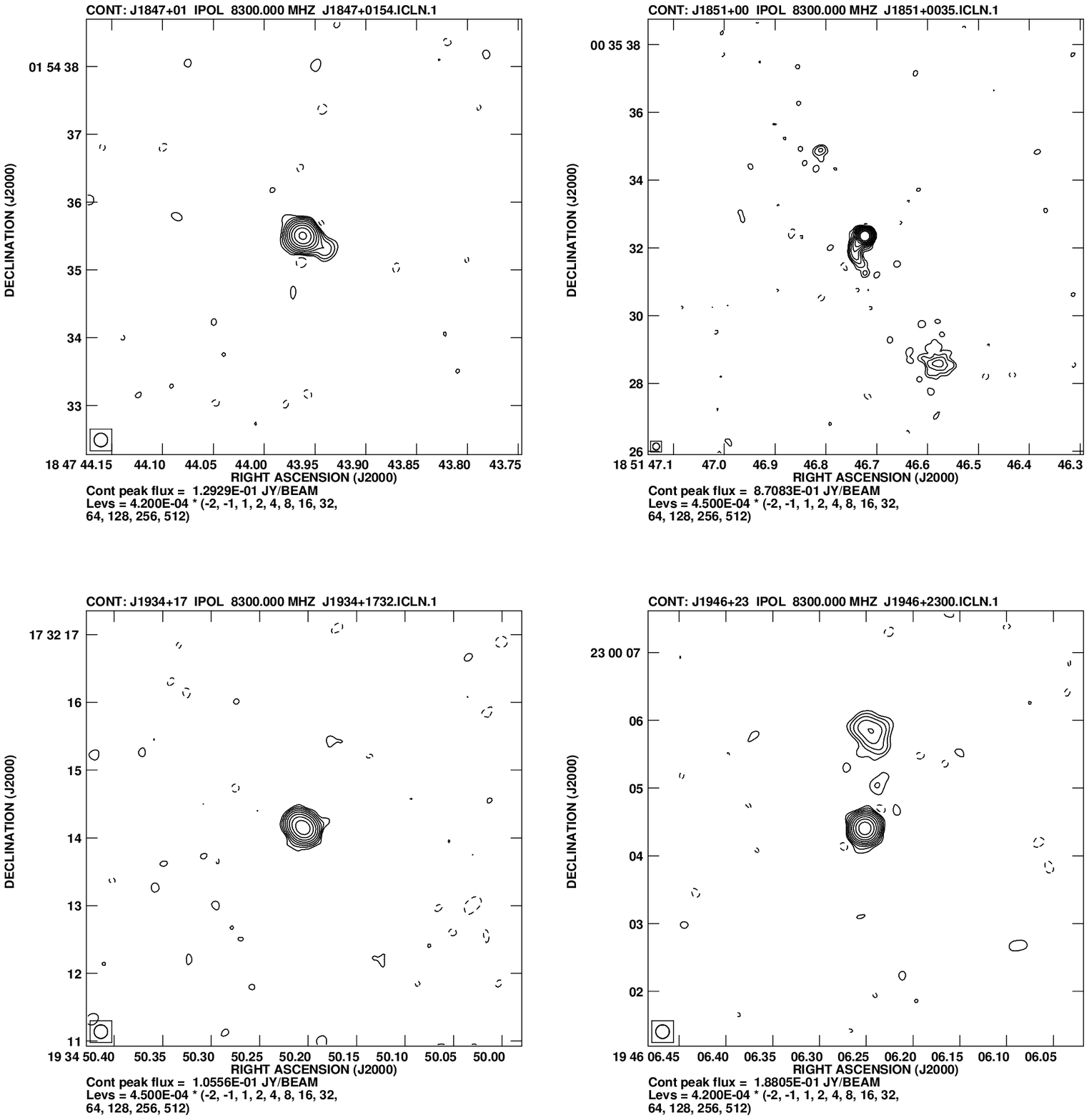}

\end{document}